\documentstyle[prl,aps,preprint,epsfig]{revtex}

\begin{document}

\title{CONTROLLING ACTIVATED PROCESSES}

\author{M.I. Dykman and Brage Golding} 

\address{Department of Physics and Astronomy, Michigan State
University, East Lansing, MI 48824, USA}

\maketitle

%\markboth{Controlling Activated Processes}{Dykman and Golding}

%\pagestyle{myheadings}
% Comment this out to remove the running heads

%\keywords{Activated escape, optical trapping, periodic driving,
%fluctuational paths}
% Keywords have to before the abstract I'm afraid.

\begin{abstract}
  The rates of activated processes, such as escape from a metastable
  state and nucleation, are exponentially sensitive to an externally
  applied field. We describe how this applies to modulation by
  high-frequency fields and illustrate it with experimental
  observations. The results may lead to selective control of
  diffusion in periodic potentials, novel control mechanisms for
  crystal growth, and new separation techniques.
  
\end{abstract}

\section{Introduction}

Activated processes are responsible for large qualitative changes in
broad classes of systems. A well-known example is
escape from a potential well, in which fluctuations carry the system
over a potential barrier. Activated escape underlies diffusion in
crystals, protein folding, and provides a paradigm for activated
chemical reactions. Another example is nucleation in phase
transitions. It would be advantageous to control the probabilities of
activated processes by applying a comparatively weak external
force. The idea is that the force need not be solely responsible for
driving the system over the barrier; it only must appropriately
influence fluctuations.

A familiar phenomenon which has elements of control of activated processes is
stochastic resonance (SR)\cite{SR1}. In SR, an adiabatic
modulation of the system parameters by a slowly varying field is
usually assumed. The strong effect of the field can be readily
understood in this case, if one notices that the probability of a
thermally activated process is $W\propto \exp(-R/k_BT)$, where $R$ is
the activation energy (the barrier height for escape from a potential
well). Even a comparatively small field-induced modulation $\delta R$
of the activation energy, $|\delta R| \ll R$, greatly affects $W$
provided $|\delta R| > k_BT$, with $\ln W$ being linear in the
modulation amplitude.

In contrast, one might expect that a high-frequency field would just
``heat up'' the system by changing its effective temperature. The rate
$W$ would then be incremented by a term proportional to the field
intensity $I$ rather than the amplitude $A\propto I^{1/2}$.  This is
indeed the case in the weak-field limit
\cite{Larkin,Devoret}. However, one may ask
what happens if the appropriately weighted field amplitude is not
small compared to the fluctuation intensity (temperature).

Recent results
\cite{Dykman-97a,Golding_PRL-99,Luchinsky-99,McCann_in_prep} show
that, counter-intuitively, for high-frequency driving the change of
$\ln W$ is linear in $A$, over a broad range of $A$. The exponential
effect of nonadiabatic driving leads to a number of new phenomena not
encountered in SR, including resonant (in the field frequency) rate
enhancement. This provides the basis for much of the selectivity and
flexibility in controlling fluctuations, as we now outline.

\section{General formulation}

One can effectively control activated processes because, although they
happen at random, the trajectories of the system in an activated process
are close to a specific trajectory. The latter is called the
optimal path for the corresponding process \cite{Freidlin-98}. The
effect of the driving field accumulates along the optimal path, giving
rise to a linear-in-the-field correction to the activation energy $R$.
It can be calculated if the fluctuational dynamics of the system is
known. Alternatively, response to the driving can be found
experimentally, and in fact used to investigate the system dynamics
away from the metastable state.

A fairly general case in which the field effect can be evaluated and
insight gained into the underlying mechanism is a dynamical system
driven by a stationary colored Gaussian noise $f(t)$ with a power
spectrum $\Phi(\omega)$ of arbitrary shape
\cite{Dykman_Springer-00}. The Langevin equation of motion is of the
form:
\begin{equation}
\label{Langevin}
\dot q = K(q;t) + f(t),\quad K(q;t) = K_0(q) + F(t), \;
F(t+\tau_F)=F(t),
\end{equation}
\noindent
where $\tau_F$ is the period of the driving force $F(t)$.

For small characteristic noise intensity $D=\max\Phi(\omega)/2$, the
system mainly performs small fluctuations about its periodic
metastable state $q_a(t)$. Large fluctuations, like those leading to
escape from the basin of attraction to $q_a$, require large bursts
of $f(t)$ which would overcome the restoring force $K$. The
probability densities of large bursts of $f(t)$ are exponentially
small, $\propto \exp[-(2D)^{-1}\int dt\,dt'\, f(t)\hat{\cal
F}(t-t')f(t')]$, and exponentially different depending on the
form of $f(t)$ [$\hat{\cal F}(t)$ is given by the Fourier transform of
$2D/\Phi(\omega)$]. Therefore for any state $q_f$ into which the
system is brought by the noise at time $t_f$, there exists a
realization $f(t)=f_{\rm opt}(t|q_f,t_f)$ which is exponentially more
probable than the others. This optimal realization and the
corresponding optimal path of the system $q_{\rm opt}(t)$
provide the minimum to the functional
\begin{eqnarray}
{\cal R}[q(t),f(t)] ={1\over 2}\int\!\!\int%_{-\infty}^{\infty} 
dt\,dt'\,f(t)\hat {\cal F}(t-t')f(t')
+ \int%_{-\infty}^{\infty} 
dt\, \lambda(t) \left[\dot q - K(q;t)-f(t)\right]
\label{varfunct}
\end{eqnarray}
\noindent 
(the integrals are taken from $-\infty$ to $\infty$). The Lagrange
multiplier $\lambda(t)$ relates $f_{\rm opt}(t)$ and $q_{\rm opt}(t)$
to each other [cf. Eq.~(\ref{Langevin}); $\lambda(t) = 0$ for
$t>t_f$].  The activation rate has the form
\begin{equation}
\label{escape_rate}
W =C \exp[-R/D], \; R = \min{\cal R}.
\end{equation}
The exponent $R$ can be obtained for an arbitrary noise spectrum and
an arbitrary periodic driving by solving the variational problem
(\ref{varfunct}) numerically, with appropriate 
boundary conditions \cite{Dykman-97a,Dykman_Springer-00}.

We now turn to the case where the driving force $F(t)$ is
comparatively weak, so that the field-induced correction $|\delta R|
\ll R$. Nonetheless, $|\delta R|$ may exceed $D$ and thus strongly
change the rate $W$ (\ref{escape_rate}).  To first order, $\delta R$
can be obtained by evaluating the term $\propto F(t)$ in
(\ref{varfunct}) along the path $q_{\rm opt}^{(0)}(t),f_{\rm
opt}^{(0)}(t)$, $\lambda^{(0)}(t)$ calculated for $F=0$.  Special care
has to be taken when activated escape and nucleation are
analyzed. Here in the absence of driving, the optimal path is an
instanton, the optimal fluctuation may occur at any time $t_c$. The
field $F(t)$ lifts the time degeneracy of escape paths. It
synchronizes optimal escape trajectories, selecting one per period, so
as to minimize the activation energy of escape $R$
\cite{Dykman-97a,Golding_PRL-99}. The correction $\delta R$ should be
evaluated along the appropriate trajectory,
\begin{eqnarray}
\label{escape_chi}
\delta R = \min_{t_c}\delta R(t_c),\quad \delta R(t_c)=
\int_{-\infty}^{\infty} dt\,
\chi(t-t_c)F(t),\quad \chi(t)=
-\lambda^{(0)}(t).
%\nonumber\\
%&&\equiv \sum_n\tilde\chi(n\omega_F)F_n\exp(in\omega_Ft_c),
\end{eqnarray}

Eq.~(\ref{escape_chi}) provides a closed-form expression for the
change of the time-averaged activation rate $\bar W$, for an arbitrary
spectrum of the driving field $F(t)$. Clearly $\ln \bar W$ is linear
in $F$, and the corresponding coefficient $\chi$ is therefore called
the logarithmic susceptibility (LS) \cite{Dykman-97a}. Because of
minimization over $t_c$, the change of $\ln \bar W$ is nonanalytic in
$F(t)$, which leads to a number of observable consequences. The LS has
been evaluated for overdamped and underdamped white-noise driven
systems\cite{Dykman-97a}. Extensive numerical and analog simulations
of the escape rate in driven systems \cite{Luchinsky-99} are in
excellent qualitative and quantitative agreement with the theory,
including the prefactor \cite{Dykman-97a,Golding_PRL-99,M&S-00}, over
a broad range of field amplitudes.

\section{Dynamical symmetry breaking in an optical trap}

A simple physical system which embodies a number of activated
phenomena is a mesoscopic dielectric Brownian particle trapped by a
strongly focused laser beam creating an optical gradient trap,
i.e. ``optical tweezers'' \cite{Ashkin-86}. Techniques based on
optical tweezers have found broad applications in contactless
manipulation of objects such as atoms, colloidal particles, and
biological materials. Activated escape can be studied using a dual
optical trap generated by two closely spaced parallel light
beams. This was used initially to investigate the
synchronization of interwell transitions by low-frequency (adiabatic)
sinusoidal forcing \cite{Simon-92}.

Quantitative characterization of activated processes requires that the
double-well confining potential of a dual trap $U({\bf r})$ be
adequately determined. The corresponding measurement, for a transparent
spherical silica particle of diameter $2R = 0.6~\mu$m optically
trapped in water, was reported recently \cite{McCann-99}.

In the experiment \cite{McCann-99}, all three coordinates of the
particle are determined simultaneously. The double-well potential
$U({\bf r})$ is found directly from the measured stationary
distribution $\rho({\bf r}) = Z^{-1}\exp[-U({\bf r})/k_BT]$. From the
Kramers theory \cite{Kramers}, it is possible then to calculate the
rates $W_{ij}$ ($i,j=1,2$) of activated transitions between the minima
of $U({\bf r})$. For the range of $U({\bf r})$ in which $W_{ij}$
vary by nearly 3 orders of magnitude, the calculated values of
$W_{ij}$ are in excellent quantitative agreement with the results of
direct measurements. This provides a direct model-free test of the
multidimensional Kramers rate theory, with no adjustable parameters.

The double-beam trap can also be used to investigate the effect of
ac-modulation on transition rates. An interesting application of this
effect is to direct the diffusion of a particle in a spatially
periodic potential \cite{Magnasco-93}. For a generic periodic potential,
the ac-induced change of the activation barrier differs depending on
the direction in which the particle moves (right or left, for
example). This makes the probabilities of transitions to the right and
to the left exponentially different and results in diffusion in the
direction of more frequent transitions.

An effect closely related to directed diffusion, but more amenable to
testing using optical trapping, is ac-field induced localization in
one of the wells of a symmetric double-well potential. Both effects
should occur if the field breaks the spatio-temporal symmetry of the
system. The ratio of the period-averaged stationary populations $\bar
w_1, \bar w_2$ of the wells is determined by the ratio of the
period-averaged transition rates $\bar W_{ij}$,
\begin{equation} 
\bar w_1/\bar w_2 = \bar W_{21}/\bar W_{12} \propto \exp(\left[\delta
R_1-\delta R_2\right]/k_BT),
\label{population_ratio}
\end{equation}
\noindent where $\delta R_{1,2}$ are the field-induced corrections
(\ref{escape_chi}) to the activation energies of escape from wells
1,2.

%\begin{figure}
\begin{center}
\epsfxsize=3.0in                %so many inches wide
\leavevmode\epsfbox{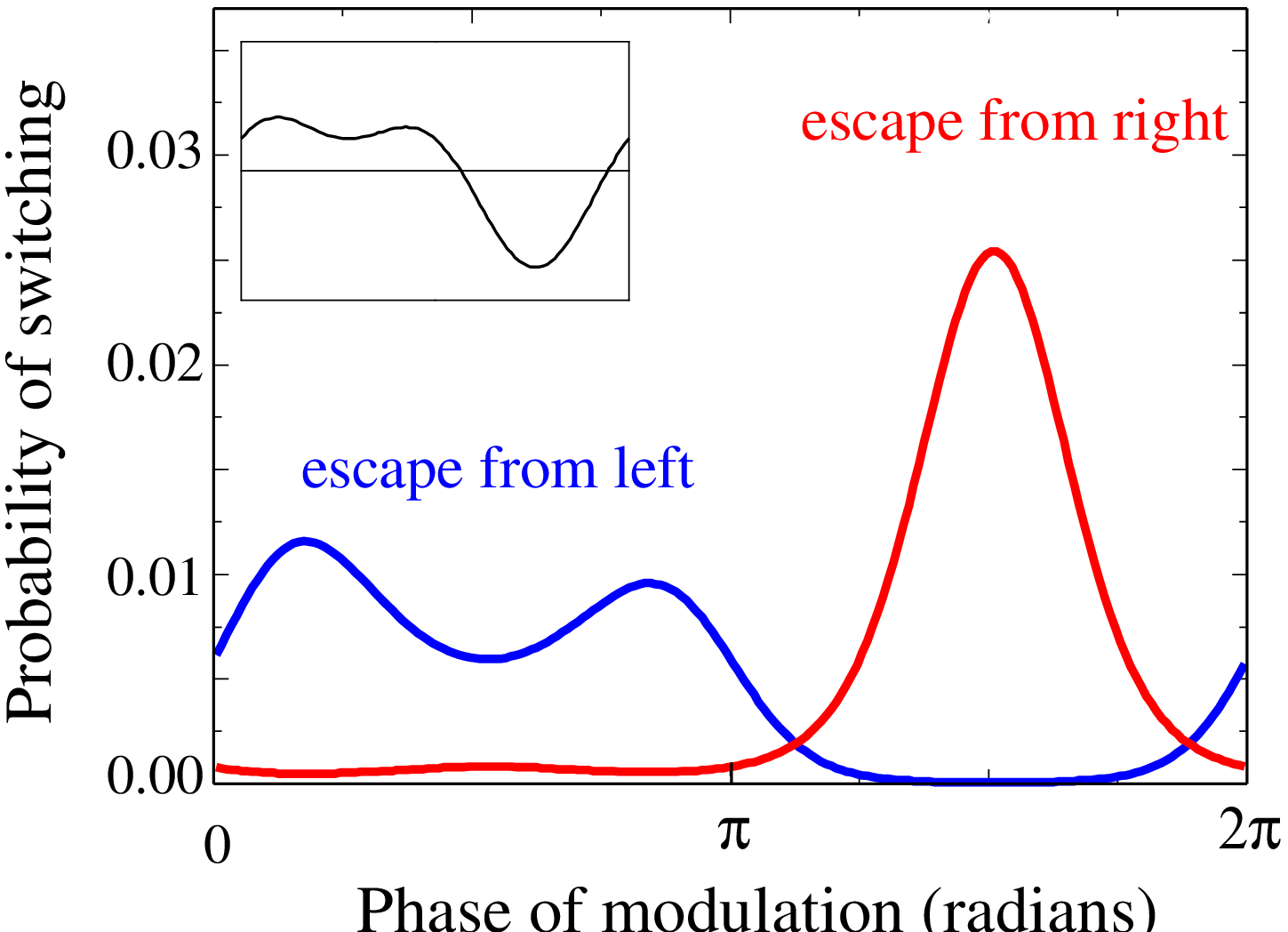}
%\caption
\end{center}

\noindent
{\footnotesize Figure 1. The least-squares fits to the experimentally determined
instantaneous time-dependent switching probabilities $W_{ij}(t)$ for a
particle in the adiabatically modulated double-beam trap, over a cycle
$\omega_F t$ of the modulating waveform.  The phase angle between the
first and second harmonics is $\phi_{12}=\pi/2$. When the phase angle
is incremented by $\pi$, the escape rates from the left and right
wells interchange. The inset shows the
instantaneous difference between the heights of the potential barriers
in the two wells.}
%\label{fig:adiabatic} 
%\end{center}
%\end{figure}

\hfill

The experimental data \cite{McCann_in_prep} on effective localization
due to the field-induced symmetry breaking are shown in
Fig.~1. The experiment is conducted with a symmetric
double-well potential and  barrier height  $ \approx 7.5\,
k_BT$. The intensity of the laser beams is then modulated so that the
well depths are changed by $\delta U_1(t) = -\delta U_2(t) = {\rm
const}\times [\sin(\omega_F t) + (1/2)\sin(2\omega_F t + \phi_{12})]$.
The modulation amplitude is $\approx 2.5\,k_BT$. The frequency
$\omega_F/2\pi$  varies between 1 and 100 Hz,
which covers the range from adiabatically slow to nonadiabatic
modulation. Over this range, field-induced re-population occurs
between the wells for a nonsinusoidal modulation waveform, so
that $\bar w_1\neq \bar w_2$.

It is the breaking of the spatio-temporal symmetry $t\to
t+\pi/\omega_F, {\bf r} \to -{\bf r}$ that leads to the escape rate
from one of the wells being on average much bigger than from the
other, as seen from Fig.~1. In turn, this leads to a
higher population in one of the wells. Not only is it observed
under slow modulation, as evidenced by Fig.~1, but a
population difference of $20\%$ is also observed far into the
nonadiabatic regime.  This is sufficient to create significant
directional diffusion, and demonstrates the onset of {\em dynamical}
symmetry breaking. The effect would not arise if nonadiabatic driving
led just to ``heating'' of the particle.

\section{Conclusions}
Investigation of methods to control activated processes is
currently at a very exciting stage. The importance of the problem and
its relevance to many areas, from condensed-matter physics to
biophysics, is becoming increasingly appreciated. The results outlined
here show that there is a fairly general approach to controlling
fluctuations, and the first experiments on overdamped systems show
that such control can indeed be exercised. A broad range of problems
remains unexplored. They include such issues as a microscopic theory
of driven many-body systems and experimental exploration of
underdamped driven systems. We also envision practical applications of
these results, starting with the development of new highly selective
colloidal separation as well as crystal growth techniques.

\section*{Acknowledgments}
We are grateful to Lowell McCann and Vadim Smelyanskiy with whom many
of the results discussed above were obtained. This research was
supported by the NSF through grants DMR-9971537 and PHY-0071059.

%\appendix % Reset the environments to Appendix style

%%%%%%%%%%%%%%%%%%%%%%%%%%%%%%
% For BiBTeX users, just uncomment the following two lines
%\bibliographystyle{unsrt}
%\bibliography{mybibfile}

\begin{thebibliography}{10}
\bibitem{SR1} For reviews on stochastic resonance see M.I.~Dykman,
D.G.~Luchinsky, R.~Mannella, P.V.E.~McClintock, N.D.~Stein, and
N.G.~Stocks, {\em  Stochastic resonance in perspective},
Nuov. Cim. D {\bf 17}, pp.~661-83 (1995); L.~Gammaitoni,
P.H.~H\"anggi, P.~Jung, and F.~Marchesoni, {\em  Stochastic
resonance}, Rev. Mod. Phys. {\bf 70}, pp.~223-88 (1998);
R.D.~Astumian and F.~Moss, {\em  Overview: the constructive role
of noise in fluctuation driven transport and stochastic resonance},
Chaos {\bf 8}, pp.~533-8 (1998); K.~Wiesenfeld and F.~Jaramillo, {\em
 Minireview of stochastic resonance}, Chaos {\bf 8},
pp.~539-48 (1998).



\bibitem{Larkin} A.I. Larkin and Yu.N.~Ovchinnikov, {\em 
Resonance reduction of the lifetime of the metastable state of tunnel
junctions}, J. Low Temp. Phys.  {\bf 63}, 317-29 (1986); B.I.~Ivlev
and V.I.~Mel'nikov, {\em  Effect of resonant pumping on
activated decay-rates}, Phys. Lett. A {\bf
116}, 427-8 (1986); S.~Linkwitz and H.~Grabert, {\em 
Enhancement of the decay rate of a metastable state by an external
driving force}, Phys. Rev. B {\bf 44} 11901-10 (1991).


\bibitem{Devoret} M.H. Devoret, %et al.,
D. Esteve, J.M. Martinis, A. Cleland, and J. Clarke, 
{\em  Resonant activation of a Brownian particle
out of a potential well: microwave-enhanced escape from the
zero-voltage state of a Josephson junction}, Phys. Rev. B {\bf 36},
58-73 (1987).

\bibitem{Dykman-97a} M.I.~Dykman, % et al.,
H.~Rabitz, V.N.~Smelyanskiy, and B.E.~Vugmeister, 
{\em Resonant directed diffusion in nonadiabatically
driven systems}, Phys. Rev. Lett. {\bf 79}, 1178--81 (1997);
%
%\bibitem{Smelyanskiy-97b} 
V.N.~Smelyanskiy, % et al.,
M.I.~Dykman, H.~Rabitz, and B.E.~Vugmeister, 
{\em Fluctuations, escape,
and nucleation in driven systems: logarithmic
susceptibility}, Phys. Rev. Lett. {\bf 79}, 3113--16 (1997).

\bibitem{Golding_PRL-99} V.N.~Smelyanskiy, M.I.~Dykman, and
B.~Golding, {\em Time oscillations of escape rates in periodically
driven systems}, Phys. Rev. Lett. {\bf 82}, 3193--3196 (1999).

\bibitem{Luchinsky-99} D.G.\ Luchinsky, % et al.,
R.~Mannella, P.V.E.\ McClintock, M.I. Dykman, and V.N. Smelyanskiy, 
{\em Thermally activated escape of
driven systems: the activation energy"}, J. Phys. A {\bf 32},
L371-7 (1999); %\bibitem{Dykman_Chaos-01}  
M.I. Dykman, B.~Golding, L.I.~McCann,
V.N.~Smelyanskiy, D.G.~Luchinsky, R.~Mannella, and P.V.E.~McClintock,
{\em Activated escape of periodically driven systems}, to be
published.


\bibitem{McCann_in_prep} L.I. McCann, M.I. Dykman, and B. Golding,
{\em Controlling diffusion with light}, in preparation.

\bibitem{Freidlin-98}
M.~I. Freidlin and A.~D. Wentzell, {\em Random Perturbations of Dynamical
Systems}, 2nd ed., Springer, New-York (1998).

\bibitem{Dykman_Springer-00} M.I. Dykman and B. Golding, {\em
Controlling large fluctuations: theory and experiment}, in Lecture
Notes in Physics {\bf 557}, eds. J.A.~Freund and T.~P\"oschel,
Springer, Berlin (2000), 365--377.

\bibitem{M&S-00} The problem of the prefactor has been also addressed
by R.S. Maier and D.L. Stein, {\em Noise-Activated Escape from a
Sloshing Potential Well}, Phys. Rev. Lett. {\bf 86} (2001).

\bibitem{Ashkin-86} A.~Ashkin, J.M.~Dziedzic, J.E.~Bjorkholm, and
S.~Chu, {\em Observation of a single-beam gradient force optical trap
for dielectric particles}, Optics Lett. {\bf 11}, 288--291 (1986).

\bibitem{Simon-92} A.~Simon A. and A.~Libchaber, {\em Escape and
synchronization of a Brownian particle},
Phys. Rev. Lett. {\bf 68}, 3375--3378 (1992)

\bibitem{Kramers} H.~Kramers, {\em Brownian motion in a field of
force and the diffusion model of chemical reactions},
Physica (Utrecht) {\bf 7}, 284--304 (1940); 
R.~Landauer and J.A.~Swanson, {\em Frequency
factors in the thermally activated process}, Phys. Rev. {\bf
121}, 1668--1674 (1961).



\bibitem{McCann-99} L.I.~McCann, M.I.~Dykman, and B.~Golding
{\em Thermally activated transitions in a bistable three-dimensional
optical trap}, Nature {\bf 402}, 785--787 (1999).

\bibitem{Magnasco-93} M.~Magnasco, {\em Forced thermal ratchets},
Phys. Rev. Lett. {\bf 71}, 1477--80 (1993); for a review see
F.~Julicher, A.~Ajdari, and J.~Prost, {\em Modeling molecular motors},
Rev. Mod. Phys. {\bf 69}, 1269-1281, (1997).


\end{thebibliography}

\end{document}